# In-plane anisotropic quantum confinement effect in ultrasmall SnS sheets

Abdus Salam Sarkar[1,2#], Anita Kumari[1,2#], Anchala[1,2], Nagaraju Nakka[1,2], Rajeev Ray[1,2], Emmanuel Stratakis[3,4] and Suman Kalyan Pal[1,2]*

[1]School of Basic Sciences, Indian Institute of Technology Mandi, Kamand, Mandi-175005, Himachal Pradesh, India.

[2]Advanced Materials Research Center, Indian Institute of Technology Mandi, Kamand, Mandi-175005, Himachal Pradesh, India.

[3]Institute of Electronic Structure and Laser, Foundation for Research and Technology Hellas, Heraklion, 700 13 Crete, Greece.

[4]Physics Department, University of Crete, Heraklion, 710 03 Crete, Greece.

*Corresponding Author: Tel.: +91 1905 267040; Fax: +91 1905 237924

E-mail: suman@iitmandi.ac.in

**Author Contributions**

[#] A. S. Sarkar and S. Kumari contributed equally.

**ABSTRACT**

Black phosphorus (BP) analogous tin(II) sulfide (SnS) has recently emerged as an attractive building block for electronic devices due to its highly anisotropic response. Two-dimensional (2D) SnS has shown to exhibit in-plane anisotropy in optical and electrical properties. However, the limitations in growing ultrasmall structures of SnS hinder the experimental exploration of anisotropic behavior in low dimension. Here, we present an elegant approach of synthesizing highly crystalline nanometer-sized SnS sheets. Ultrasmall SnS exhibits two distinct valleys along armchair and zig-zag directions due to in-plane structural anisotropy like bulk SnS. We show that in such SnS nanosheet dots, the band gaps corresponding to two valleys are increased due to quantum confinement effect. We particularly observe that SnS quantum dots (QDs) show excitation energy dependent photoluminescence (PL), which originates from the two non-degenerate valleys. Our work may open up an avenue to show the potential of SnS QDs for new functionalities in electronics and optoelectronics.

## I. INTRODUCTION

Emerging two dimensional (2D) materials and their quantum dots (QDs) are promising platform for realizing tunable quantum effect for next generation quantum electronic, optoelectronic, photonic and sensing applications[1-3]. In particular, group IV$_A$-VI metal monochalcogenides (MMCs), exhibiting lower crystal symmetry ($C_{2v}$, broken inversion) orthorhombic structures with chemical formula MX (M=Sn/Ge and X=S/Se) have attracted significant attention for their unique in-plane anisotropy in intrinsic physical properties[4-10]. Among all MXs, layered tin(II) monosulfide (SnS)[7, 8, 11-16] exhibiting the puckered crystal structure (Fig. 1a) and broken inversion symmetry gives rise to the highest in-plane anisotropy along its armchair (*x*) and zig-zag (*y*) directions. This anisotropy has detected in Raman response[8, 17], absorption[7, 12], photoluminescence[7, 12], second harmonic generation[8], ferroelectricity[8, 13], and mutiferoelecricity. Beyond this, the anisotropy in electronic [11, 13-15] and optical[7, 12] responses have been visualized, and efforts were made to exploit such anisotropic properties in electronic, valleytronic and neuromorphic device applications[7, 18, 19]. Most importantly, all these fascinating anisotropic behaviors of SnS appear at room temperature.

So far, ultrathin- and few- layers of SnS sheets has been prepared using mechanical[12, 20, 21], physical[8, 13, 22], and chemical methods[14, 15, 23-27]. O'Brien and co-workers[23] have isolated 3-4 bilayers SnS with 50-100 nm lateral dimensions. We have successfully isolated bi-layer (1.10 nm) SnS nanosheets of 170 nm lateral dimension[24]. However, the lateral dimensions of such sheets were much larger than the excitonic Bohr radius of SnS ($r_B$~7 nm)[28]. Particularly, when the lateral dimension of an ultrathin 2D nanosheet falls below the excitonic Bohr radius, the quantum-size effects[29-31] and edge effects[32-36] become significant, resulting in fascinating quantum phenomena in physical and optical properties at their quantum limits. The photoluminescence (PL) intensity enhancement due to quantum confinement or size effects across a few layer 2D planes have been investigated[30, 31]. In particular, the edge-effect in QDs of few-layer 2D materials gives rise to a significant shift in the PL peak position and intensity [35-37]. In the first report, Tilley and co-workers[38] presented the preparation of SnS nanoparticles (NP) of 6 nm in size. Further, size controlled QDs have been achieved by Han et al. [39]. Klinke and co-workers[40] have focused on the preparation of SnS nanostructures of different sizes and shapes. Huang et al.[41] used a mechanical method to synthesize bare SnS nanocrystals. Wang and co-workers[42] prepared SnS QDs of different sizes (~7 and ~4 nm). However, such bottom-up approaches suffer from obtaining QDs of high crystalline quality. On the contrary, top-down approaches were widely used to produce high quality QDs of graphene and other 2D materials[43-45]. However, the preparation of highly crystalline electronic grade, ultra-small and few-layered SnS QDs has not been realized yet.

It has been reported that 2D monolayer of SnS exhibits two distinct valleys along its anisotropic armchair ($x$) and zigzag ($y$) directions leading to two non-degenerate band gaps [46]. In fact, these two distinct valleys can be directly accessible at room temperature in bulk SnS , identified at 1.24 and 1.48 eV respectively [7]. Nonetheless, Chen et al. [12] observed two nearly non-degenerate band gaps along the $\varGamma X$ (armchair) and $\varGamma Y$ (Zig-zag) directions in SnS. Nevertheless, the revelation of anisotropic in-plane quantum confinement effect in PL properties of SnS QDs at room temperature remains elusive.

In this letter, we report room temperature quantum confinement effect in anisotropic thin-layered SnS QDs. We present a novel preparation method for SnS QDs, which exhibit excitation energy dependent blue PL at ca. 2.66 eV. Our spectroscopic studies reveal in-plane anisotropy in the QDs optical properties along its armchair and zig-zag directions. It is envisaged that such fascinating optical properties could find applications in quantum photonic and optoelectronic devices.

## II. EXPERIMENTAL

### A. Preparation of SnS NSs and QDs.

SnS quantum dots were synthesized from SnS NSs obtained from bulk SnS (99.999%, Sigma Aldrich, USA). The bulk SnS powder was dispersed in acetone with the help of bath ultrasonication for 10 h. Then SnS dispersion was centrifuge at 8000 rpm for 15 min to isolate the thin layer (supernatant) from its bulk counterpart. A basic solution of 0.1 M NaOH was prepared by dissolving 0.02 g of solid NaOH in 5 ml of DI water. The $_{pH}$ of the solution was maintained at 13.7. The supernatant containing SnS nanosheets were mixed with NaOH solution in equal volume. After mixing the yellow-colored solution became colorless. The mixed solution was kept for ultrasonication in a bath sonicator (LMUC-4 from Spectrochrome Instruments, India) for 5 h under 100 W power and 40 kHz ultrasonic frequency. The temperature was maintained at less than 50 $^{o}$C throughout the ultrasonication. The transparent solution obtained after ultrasonication was centrifuged at 8000 rpm for 15 minutes to isolate the small-sized SnS. The supernatant was collected and used for further experiments.

### B. Microscopic characterization.

High resolution TEM (HRTEM) analysis of the samples was performed by a TECNAI G2 200 kV (FEI, Electron Optics) electron microscope with 200 kV input voltage. Synthesized SnS sample was drop casted onto 300 square mesh copper grids (from ICON) covered with a carbon film. The drop casted grids were dried at room temperature. Further, precautions were taken to avoid dust particles

Tapping mode AFM images were collected in an instrument from Dimension Icon with ScanAsyst, Bruker, USA. The sample was prepared by drop casting dispersion of as-prepared SnS solution on cleaned $SiO_2$ substrate. The drop casted sample was dried and placed on the AFM stage to scan. Pertinent scanning parameters were as follows: scan rate for measurement- 0.628 Hz; aspect ratio- 1:1; resolution- 256 samples/line, 256 lines.

C. **Spectroscopic measurements.**

The UV-vis extinction spectrum was recorded using Shimadzu UV-2450 spectrometer (Agilent Technologies, USA). All measurements were carried out with dispersion of SnS nanosheets and QDs in solvent using a quartz cuvette (path length of 10 mm) at room temperature. The PL and PLE spectra were recorded on a Cary Eclipse PL spectrophotometer (Agilent Technologies) at room temperature.

The PL decay kinetics were recorded on a time correlated single photon counting (TCSPC) setup from Horiba Scientific Delta Flex TCSPC system with pulsed LED source having light photon energy 2.77 eV (447 nm). Ludox has been used to calculate IRF for de-convolution of PL kinetics of the sample.

Raman spectroscopic measurements were carried out in the backscattering geometry using a confocal microRaman spectrometer (LABRAM HR Evolution, Horiba JobinYvon SAS) coupled with a Peltier cooled CCD and spectrometer. The Raman spectra were collected with 532 nm laser excitation focused using a 100X long-working distance objective and 1800 g/mm grating, after calibrating the spectrograph with Si substrate (the peak located at 520 cm$^{-1}$ was taken as an internal reference). The incident laser power density was 3.21 mW/µm$^2$. The samples were prepared by drop casting of SnS nanosheets and QDs on cleaned Si substrates and drying them inside the glove box ($O_2$ < 0.5 ppm and $H_2O$ < 0.5 ppm) at room temperature.

The X-ray diffraction pattern of SnS samples was recorded by an X-Ray diffractometer from Bruker D8 Advance, Twin-Twin technology with a wavelength of 1.54 Å (Copper source). The SnS QD solution was deposited on the glass substrate, which was dried at room temperature.

### III. RESULTS AND DISCUSSION

SnS QDs were prepared by liquid phase exfoliation (LPE) of bulk SnS crystals, followed by a modified chemical *nano-scissoring* method. The process is shown schematically in Fig. 1c. In order to investigate the crystal structure (Fig 1a) of as prepared SnS QDs, low- and high-resolution transmission electron microscopy (HRTEM) images were captured, presented in Fig. 2a and b.

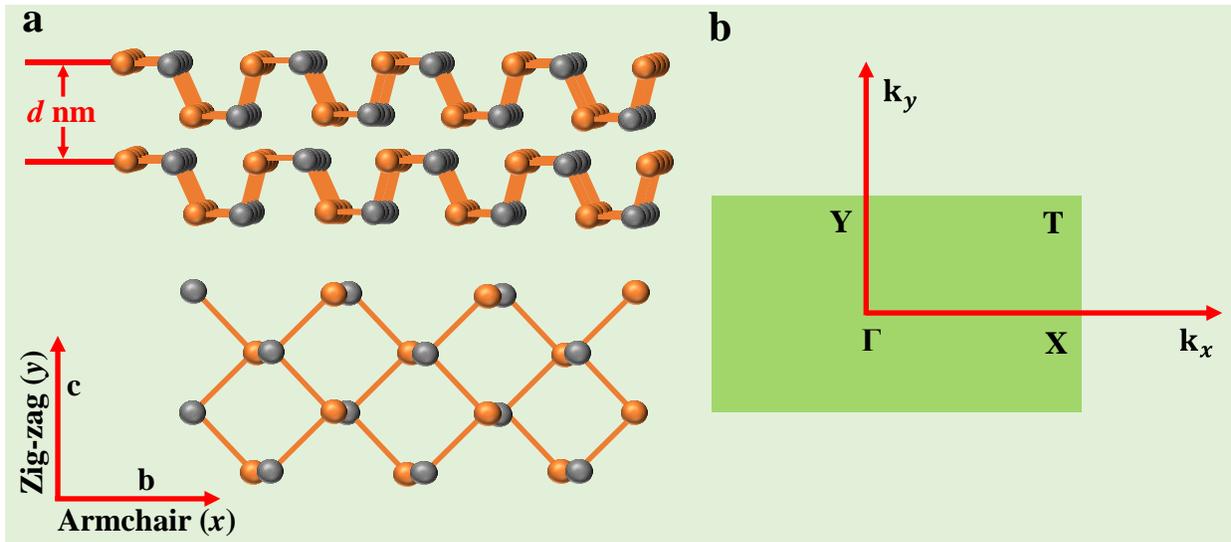

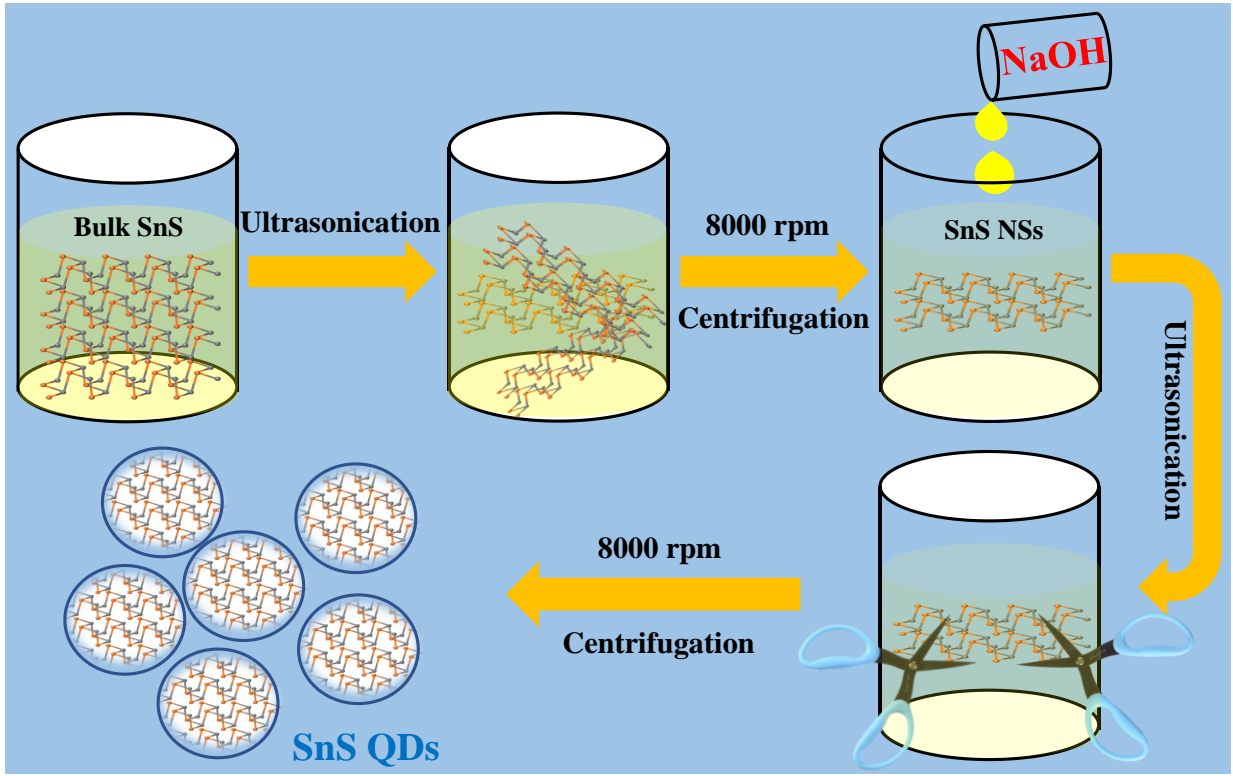

FIG. 1. Crystal structure of SnS and synthesis process of SnS QDs. **a** Side view (top panel) and top view (bottom panel). Wavy or buckled structure along armchair ($x$) and zig-zag ($y$) direction. Orange atoms are tin (Sn) and gray atoms are sulphur (S). $d$ is the thickness of each monolayer of SnS. **b** The first Brillouin zone with orthorhombic crystal symmetry points Γ, X, T, and Y. $k_x$ and $k_y$ are momentum along armchair and zig-zag directions, respectively. **c** Schematic representation of the synthesis process of SnS QDs via liquid exfoliation followed by modified *chemical nano-scissoring* treatment.

The size distribution fitted by a Gaussian curve is shown in inset of Fig. 2a, affirming the monodispersed distribution with an average size of 2.48±0.03 nm. The lattice planes are clearly visible in the HRTEM images (Supplemental material Fig. S1a)[47]. The relatively fuzzy selected area electron diffraction (SAED) pattern is appeared (Supplemental material Fig. S1b)[47] due to the ultrasmall size of the SnS QDs[48]. The lattice spacing of 0.283 nm corresponding to (040) plane is confirmed by first Fourier transform (FFT) of HRTEM images (Fig. 2b). These observations infer that the highly crystalline orthorhombic structure of bulk SnS is preserved in QDs. Moreover, SnS bulk and QDs were probed with X-Ray diffractometer to identify their crystallographic structures. Both spectra are featured with multiple diffraction peaks (Supplemental material Fig. S2a)[47], corresponding to different crystallographic planes. Among them, the peak appeared at 32º is associated with the orthorhombic (040) plane. The same orthorhombic crystal plane is apparent in SnS QDs. Further, the orthorhombic crystal structure of both SnS bulk and QDs are reconfirmed by matching XRD pattern with the PDF card no 39-0054[26]. The atomic force microscopy images of SnS QDs are acquired to quantify the vertical dimensionality (Fig. 2c). The height of QDs is varied from 1.5 nm to 3.5 nm with an average height of 2.7 nm (Fig. 2d) suggesting the presence of few (six) monolayers[24]. In order to visualize the in-plane anisotropic crystallographic orientations along armchair and zig-zag directions of a SnS QD, we exploited the capability of AFM. The atomic potential modulate the friction forces between tip and sample, which reveals high symmetry anisotropic orthorhombic lattice planes in a SnS QD (Supplemental material Fig. S2b, c)[47]. In the First Fourier transformed image, the crystal planes are clearly visible along its armchair and zig-zag directions (Fig 2e) suggesting formation of high-quality crystals of SnS. The presence of symmetry in anisotropic crystal orientations (Supplemental material Fig. S2c)[47] infers that there is orthorhombic in-plane anisotropic phase integrity in SnS QDs.

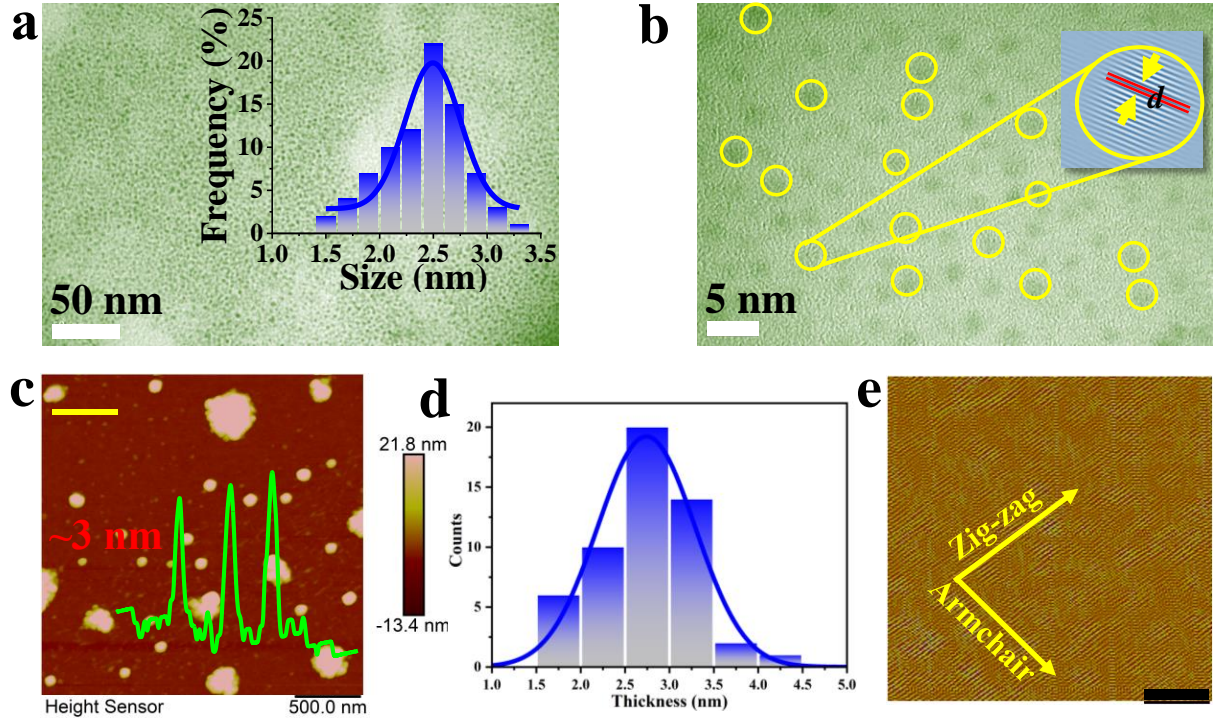

FIG. 2. Microscopic and spectroscopic characterizations of SnS QDs. **a** Low-resolution TEM images (Inset: Histogram of lateral size distribution of SnS QDs with an average size ~2.48 nm. Size distribution is obtained by monitoring 85 QDs. The blue solid line is the result of Gaussian fitting). **b** High-resolution TEM image (Inset: lattice spacing). **c** Atomic force microscopic image (Inset: Height profile of SnS QDs measured along the yellow line). Scale bar is 500 nm. White objects are the clusters of SnS QDs. **d** Histogram of thickness distribution which is obtained from 56 QDs. The blue solid line is the Gaussian fitting. Average thickness is ~2.75 nm. **e** Crystal lattice resolution in SnS QDs revealing the anisotropic crystallographic planes along armchair and zig-zag directions. Scale bar is 2.25 nm.

Orthorhombic puckered 2D SnS possesses $D_{2h}$ crystal symmetry (Pnma). In such crystals, 24 phonon modes are present at the center of Brillouin zone (Fig. 1b). Their phonon vibrational modes can be expressed as

$$\Gamma = 4A_g + 2B_{1g} + 4B_{2g} + 2B_{3g} + 2A_u + 4B_{1u} + 2B_{2u} + 4B_{3u} \qquad (1)$$

where, $A_g$, $B_{1g}$, $B_{2g}$, and $B_{3g}$ are Raman active modes[24, 49, 50]. **Fig. 3a** displays room temperature Raman spectra of exfoliated SnS nanosheets and QDs measured using 532 nm (2.33 eV) laser. Unlike previous studies, three characteristics Raman modes are appeared in the Raman spectrum of SnS nanosheets. The Raman peaks at ~161, 189 and 220 cm$^{-1}$ correspond to $B_{3g}$, $A_g$ (1), and $A_g$ (2) modes, respectively. These Raman modes also appear in SnS QDs. However, in SnS QDs the peak position corresponding to $A_g$ (1) mode remains the same, while $A_g$ (2) mode has red-shifted (by 8 cm$^{-1}$). The appearance of $B_{3g}$ mode indicates that SnS QDs still exhibit unique puckered orthorhombic layered structure. On contrary, the red-shift in Raman modes is attributed to the change of van der Waals interlayer interaction due to low thickness and small size, which is well agreed with other anisotropic 2D material QDs[51-53].

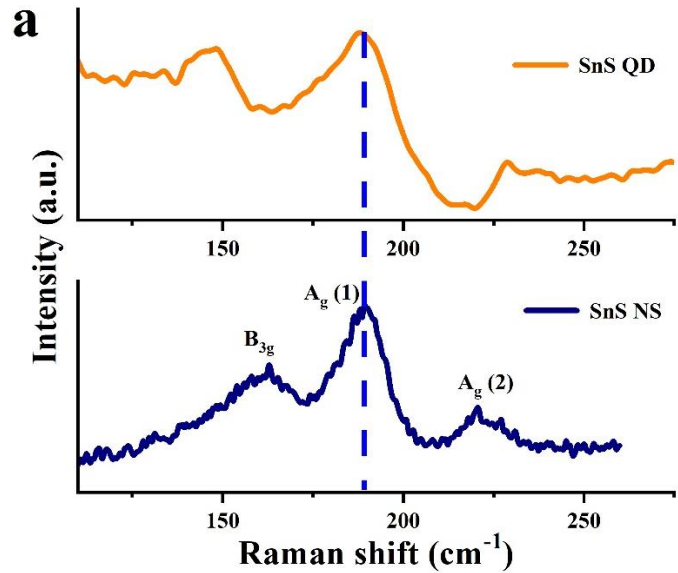

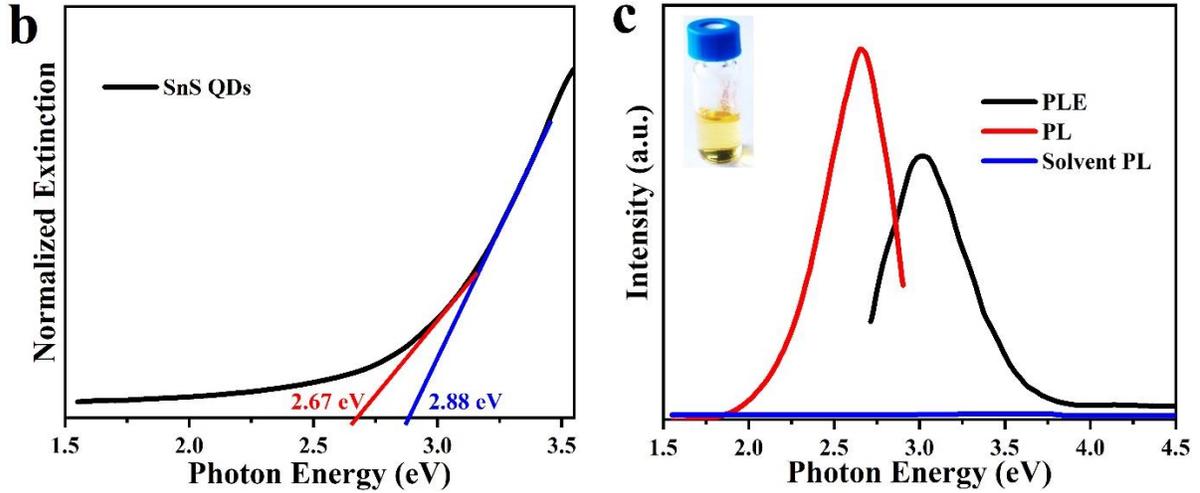

FIG. 3. Steady state optical properties of SnS QDs. **a** Room temperature Raman spectra of SnS nanosheets (NSs) and QDs. **b** Extinction spectrum of SnS QDs. Two band gaps are observed at 2.88 and 2.67 eV, respectively. Difference between two band gap energies is ~0.21 eV. **c** Photoluminescence (PL) and photoluminescence excitation (PLE) spectra (Inset: optical image of SnS QD solution) of Sns QDs.

Figure 3b shows room temperature extinction spectrum of SnS QDs. There are two distinct humps in the extinction spectrum at different photon energies, which infer the presence of two distinct band gaps in SnS QDs. These band gap energies have been extracted from the extinction spectrum and found to be 2.88 and 2.67 eV. It has been previously reported that, SnS has one pair of band gaps ($E_{g\Gamma X/a}$ and $E_{g\Gamma Y/z}$) along in-plane orthogonal high symmetry axes i.e., along puckered armchair (*x*) and zig-zag (*y*) directions[7, 46]. According to previous reports, the difference ($\Delta E_g = E_{g\Gamma X/a} - E_{g\Gamma Y/z}$) between the band gap energies in SnS is about 0.2 eV. However, the obtained energy difference ($\Delta E_g$) between the two absorption bands of SnS QDs is ~0.21 eV. Very similar energy difference between the two band positions in bulk and QDs infers the presence of anisotropic optical band dispersion at *k* point along the $\Gamma X$ and $\Gamma Y$ directions in SnS QDs like its bulk counterpart (Fig. 1b).

It is well known that the band gap of a semiconductor increases when its size approaches the excitonic Bohr radius. Considering the effective mass approximation, the size dependent band gap can be obtained from the following equations[30, 54, 55],

$$E = E_g + \frac{h^2}{8\mu r^2} - \frac{1.8e^2}{4\pi\varepsilon_0\varepsilon r} \qquad (2)$$

$$\mu = \frac{m_e m_h}{m_e + m_h} \qquad (3)$$

where, $E_g$ is the bulk band gap, $\mu$ is the reduced mass of the exciton, r is the radius of the QDs, h is the Plank's constant, e is the electronic charge and $\varepsilon$ is the dielectric constant. For SnS, $E_g$ is 1.28 eV and the effective mass of electron and hole is $0.20m_0$ and $0.16m_0$, respectively[56] where, $m_0$ is the free-electron mass. The excitonic Bohr radius,

$$R = \varepsilon a_0 \left(\frac{m_0}{\mu}\right) \qquad (4)$$

Considering $a_0$=0.053 nm and $\varepsilon$=13 for SnS[56], the excitonic Bohr radius of SnS obtained from Eq 4 is ~4.3 nm. Thus obtained excitonic Bohr radius is much larger than the experimentally obtained radius (1.25 nm) (inset of Fig. 2a) suggesting the possibility of quantum confinement effect in synthesized SnS dots. The band gap calculated from the effective mass approximation (Eq. 2) is 2.67 eV. The estimated band gap energy matches well with the experimentally measured value (from extinction spectra), which further indicates strong carrier confinement in our synthesized SnS structures.

The room temperature photoluminescence (PL) spectrum of SnS QDs dispersed in acetone is presented in Fig. 3c. The PL spectrum exhibit a strong visible emission, which is peaked at 2.66 eV (466 nm) under an excitation energy of 3.02 eV (410 nm). In contrast, no reliable PL signal was detected from acetone under the same experimental condition (Fig. 3c). In order to provide a comprehensive view of PL characteristics of SnS QDs, the PL spectrum was measured at different excitation energies (Fig. 4a). PL intensity of SnS QDs increases with decreasing excitation energy, reaches maximum at an excitation energy of 3.02 eV (410 nm), and then decreases (Fig. 4b). However, the PL peak is red-shifted with the increase in the excitation energy (Fig. 4a). The energy of the emission maximum is shifted from 2.67 (464 nm) to 2.27 (546 nm) eV as the excitation energy changes from 3.26 (380 nm) to 2.48 (500) eV, respectively (Supplemental material Fig. S3)[47]. These observations suggest that SnS QDs exhibit a typical excitation energy dependent PL behavior [30, 31, 37, 57]. Noticeably, the full width at half maximum (FWHM) of PL spectrum increases as the energy of excitation light decreases. Such variation in FWHM is indicating involvement of different luminescence processes. The dependence of PL on excitation energy has been observed for graphene and TMD QDs[31, 37, 57, 58]. In such systems, quantum confinement effect (also known as size effect)[30], surface traps[37, 59] and edge states[34, 36, 37] are the most comprehensively accepted causes for excitation dependent PL.

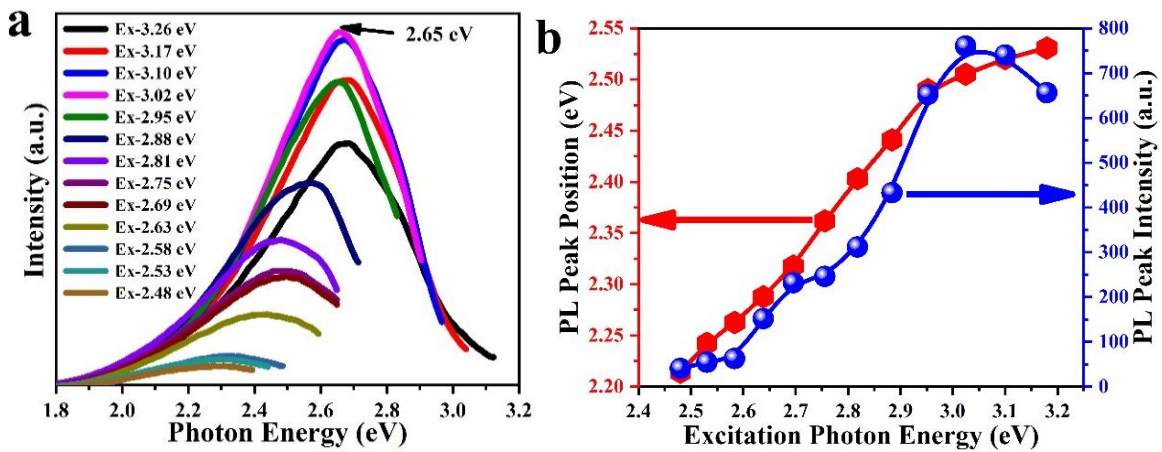

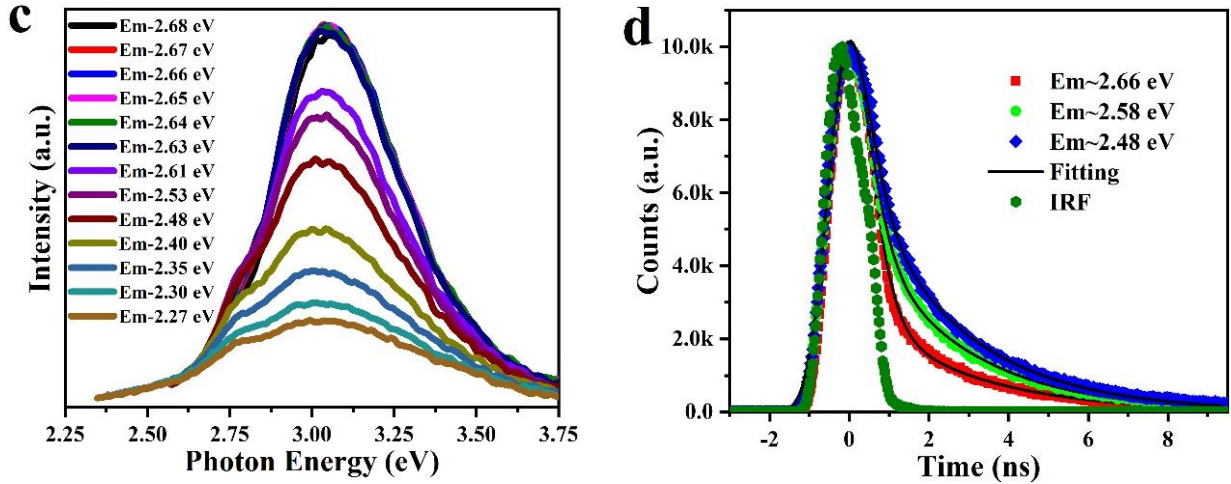

FIG. 4. Steady state and time-resolved PL properties of SnS QDs. **a** Excitation energy dependent PL spectra. The excitation energy changes from 3.18 to 2.48 eV (wavelengths from 380 to 500 nm). **b** Variation of PL peak position and intensity with excitation energy. **c** PLE spectra detected at 2.68 eV (461 nm) to 2.27 eV (546 nm). **d** PL decay kinetics of SnS QDs. The excitation photon energy is 2.77 eV (447 nm) and monitoring PL energies are 2.66, 2.58 and 2.48 eV.

To gain insight into the origin of intensity dependent PL emission, PL excitation (PLE) spectra of SnS QDs has been recorded (Fig 4c). The PLE spectra were measured by monitoring at different PL energies starting from 2.27 to 2.68 eV. There are two distinct peaks appeared in PLE spectrum. The intense peak is observed at ~3.04 eV (407 nm), while a weak shoulder is seen at 2.76 eV (448 nm). These peak positions are consistent with the absorption band edges at ~2.88 and ~2.67 eV of SnS QDs (Fig. 3b). The peak positions in PLE spectrum do not change with monitoring PL energies (Supplemental material Fig. S4)[47]. These observations infer that PL of a QD originates from two distinct states associated with its anisotropic zig-zag and armchair crystal orientation related band edges. It is worth mentioning that in case of size dependent PL characteristics of QDs, peak positions of PLE spectrum vary with monitoring PL energy. In SnS

QDs, as the PLE characteristic is independent of PL monitoring energy, excitation intensity dependent PL behavior does not depend on the size of QDs [30, 55].

To further shed light on the origin of the PL of SnS QDs, time-resolved PL (TRPL) measurements were carried out. PL decay kinetics of SnS QDs are shown in Fig. 4d. The PL decay traces were acquired in a time corelated single photon counting (TCSPC) system at PL energies 2.66 (465 nm), 2.58 (480 nm) and 2.48 eV (500 eV) following excitation at 2.77 eV (447 nm). The slope of the PL decay profile changes with varying monitoring energy. The PL decay kinetics can be fitted using a bi-exponential decay function expressed as[60, 61],

$$I(t) = A_1 \exp(-t/\tau_1) + A_2 \exp(-t/\tau_2) \tag{5}$$

where, $A_1$ and $A_2$ are the weight factors, and $\tau_1$ and $\tau_2$ are the corresponding lifetimes. It is clear from the figure that experimental data are fitted well with eq. 5. The lifetime components and corresponding weight factors obtained from fitting are listed in Table 1. At the monitoring PL energy 2.66 eV, the lifetime vales are 0.37 and 3.0 ns. These lifetime components are insignificantly varying at monitoring PL energies 2.58 and 2.48 eV. The two decay components in PL kinetics could arise due to the emission from the two separate emitting states in SnS QDs. Nonetheless, the fractional contribution of each lifetime component varies with monitoring PL energies. In particular, the fractional contribution ($f_1$) of first lifetime component ($\tau_1$) has significantly increased with lowering the monitoring PL energies *i.e.* moving towards red part of the spectrum. On contrary, the fractional contribution of second lifetime component has reduced with shifting monitoring wavelength towards red. These observations suggest that the lifetimes $\tau_1$ and $\tau_2$ are associated with PL from low and high energy emitting sates, respectively. In the previous section, we mentioned that the PL in SnS QDs originates from the valleys along armchair ($\Gamma X$) and zig-zag ($\Gamma Y$) directions. As the band gap of $\Gamma X$ valley is less, $\tau_1$ could be associated to the $\Gamma X$ valley, but $\tau_2$ is the lifetime of the PL emitted from the $\Gamma Y$ valley having high band gap.

Table 1. PL Lifetime components of SnS QDs excited at 2.77 eV.

| Monitoring PL energy (eV) | $\tau_1$ (ns) | $f_1$ (%) | $\tau_2$ (ns) | $f_2$ (%) |
|---|---|---|---|---|
| 2.66 | 0.37 | 15 | 3.0 | 85 |
| 2.58 | 0.40 | 24 | 3.1 | 74 |
| 2.48 | 0.39 | 32 | 3.0 | 68 |

## IV. CONCLUSIONS

In summary, we have synthesized SnS nanosheet-shaped QDs via a liquid-phase chemical nano-scissoring method. Our morphological studies reveal that the structural anisotropy of bulk SnS remains intact in QDs leading to two non-degenerate valleys along zig-zag and armchair directions. The band gap energies at these valleys are enhanced in QDs due to quantum confinement effect. The as-prepared QDs exhibited remarkable excitation dependent PL due to emission from the two different valleys. Our work may offer a versatile approach to prepare QDs of group IV$_A$-VI metal monochalcogenides on a large scale.

**ACKNOWLEDGEMENTS**

This work was supported by Department of Science and Technology (DST), India (DST/INT/SWD/VR/P-07/2019). A. Kumari is also acknowledging to the Department of Science and Technology, Govt. of India for the INSPIRE fellowship.

**Competing interests**

The authors declare no competing financial interest.

---

[1]   X. Liu, and M. C. Hersam, Nat. Rev. Mater. **4**, 669 (2019).

[2]   W. Han, Y. Otani, and S. Maekawa, npj Quantum Mater. **3**, 27 (2018).

[3]   A. Reserbat Plantey *et al.*, ACS Photon. **8**, 85 (2021).

[4]   F. Xia *et al.*, Nat. Rev. Phys. **1**, 306 (2019).

[5]   L. C. Gomes, and A. Carvalho, Phys. Rev. B **92**, 085406 (2015).

[6]   X. Zhou *et al.*, Adv. Sci. **3**, 1600177 (2016).

[7]   S. Lin *et al.*, Nat. Commun. **9**, 1455 (2018).

[8]   N. Higashitarumizu *et al.*, Nat. Commun. **11**, 2428 (2020).

[9]   S. Barraza Lopez *et al.*, RMP **93**, 011001 (2021).

[10]  A. S. Sarkar, and E. Stratakis, Adv. Sci. **7**, 2001655 (2020).

[11]  Z. Tian *et al.*, ACS Nano **11**, 2219 (2017).

[12]  C. Chen *et al.*, ACS Photon. **5**, 3814 (2018).

[13]  Y. Bao *et al.*, Nano Lett. **19**, 5109 (2019).

[14]  F. Li *et al.*, J. Phys. Chem. Lett. **10**, 993 (2019).

[15]  M. M. Ramin Moayed *et al.*, Nanoscale **12**, 6256 (2020).

[16]  K. C. Kwon *et al.*, ACS Nano **14**, 7628 (2020).

[17]  J. Xia *et al.*, Nanoscale **8**, 2063 (2016).

[18]  V. K. Sangwan, and M. C. Hersam, Nat. Nanotechnol. **15**, 517 (2020).

[19] Z. Lv *et al.*, Chem. Rev. **120**, 3941 (2020).

[20] N. Higashitarumizu *et al.*, MRS Adv. **3**, 2809 (2018).

[21] N. Higashitarumizu *et al.*, Nanoscale **10**, 22474 (2018).

[22] M. Patel, J. Kim, and Y. K. Kim, Adv. Funct. Mater. **28**, 1804737 (2018).

[23] J. R. Brent *et al.*, J. Am. Chem. Soc. **137**, 12689 (2015).

[24] A. S. Sarkar *et al.*, npj 2D Mater. Appl. **4**, 1 (2020).

[25] A. S. Sarkar, and E. Stratakis, J Colloid Interface Sci. **594**, 334 (2021).

[26] H. Khan *et al.*, Nat. Commun. **11**, 3449 (2020).

[27] V. Krishnamurthi *et al.*, Adv. Mater. **32**, 2004247 (2020).

[28] D. S. Koktysh, J. R. McBride, and S. J. Rosenthal, Nanoscale Res. Lett. **2**, 144 (2007).

[29] J. P. Wilcoxon, and G. A. Samara, Phys. Rev. B **51**, 7299 (1995).

[30] Z. X. Gan *et al.*, Appl. Phys. Lett. **106**, 233113 (2015).

[31] C. Y. Luan *et al.*, Appl. Phys. Lett. **111**, 073105 (2017).

[32] O. I. Micic *et al.*, Appl. Phys. Lett. **68**, 3150 (1996).

[33] D. Pan *et al.*, Adv. Mater. **22**, 734 (2010).

[34] M. Hassan *et al.*, Nanoscale **6**, 11988 (2014).

[35] Y. Li *et al.*, J. Phys. Chem. C **119**, 24950 (2015).

[36] J. Tang *et al.*, Nanoscale Res. Lett. **14**, 241 (2019).

[37] Z. Gan, H. Xu, and Y. Hao, Nanoscale **8**, 7794 (2016).

[38] Y. Xu *et al.*, J. Am. Chem. Soc. **131**, 15990 (2009).

[39] J. Han *et al.*, Small **13**, 1700953 (2017).

[40] F. Li *et al.*, J. Mater. Chem. C **6**, 9410 (2018).

[41] X. Huang *et al.*, Sci. Rep. **7**, 16531 (2017).

[42] Y. Li *et al.*, ACS Appl. Energy Mater. **2**, 3822 (2019).

[43] S. C. Dhanabalan *et al.*, Adv. Opt. Mater. **5**, 1700257 (2017).

[44] Y. Yan *et al.*, Adv. Mater. **31**, 1808283 (2019).

[45] W. Yin *et al.*, Chem. Mater. **32**, 4409 (2020).

[46] A. S. Rodin *et al.*, Phys. Rev. B **93**, 045431 (2016).

[47] See Supplemental Material at http://link.aps.org/supplemental/.

[48] L. Long *et al.*, Small **14**, 1803132 (2018).

[49] S. Zhang *et al.*, Chem. Soc. Rev. **47**, 3217 (2018).

[50] M. Park *et al.*, Sci. Rep. **9**, 19826 (2019).

[51] Y. W. Wang *et al.*, Nanoscale **9**, 4683 (2017).

[52] Z.-L. Xu *et al.*, Nat. Commun. **9**, 4164 (2018).

[53] A. S. Sarkar, and S. K. Pal, ACS Omega **2**, 4333 (2017).

[54] F. Parsapour *et al.*, J. Chem. Phys. **104**, 4978 (1996).

[55] X. L. Wu *et al.*, Phys. Rev. Lett. **94**, 026102 (2005).

[56] J. Vidal *et al.*, Appl. Phys. Lett. **100**, 032104 (2012).

[57] S. P. Caigas *et al.*, Appl. Phys. Lett. **112**, 092106 (2018).

[58] S. K. Pal, Carbon **88**, 86 (2015).

[59] Z. Gan *et al.*, Adv. Opt. Mater. **1**, 926 (2013).

[60] X. Liu *et al.*, J. Phys. Chem. C **117**, 10716 (2013).

[61] A. S. Sarkar, and S. K. Pal, J. Phys. Chem. C **121**, 21945 (2017).

# Supplemental material

## In-plane anisotropic quantum confinement effect in ultrasmall SnS sheets

Abdus Salam Sarkar[1,2,#], Anita Kumari[1,2,#], Anchala[1,2], Nagaraju Nakka[1,2], Rajeev Ray,[1,2] Emmanuel Stratakis[3,4] and Suman Kalyan Pal[1,2]∗

[1]School of Basic Sciences, Indian Institute of Technology Mandi, Kamand, Mandi-175005, Himachal Pradesh, India.

[2]Advanced Materials Research Center, Indian Institute of Technology Mandi, Kamand, Mandi-175005, Himachal Pradesh, India.

[3]Institute of Electronic Structure and Laser, Foundation for Research and Technology Hellas, Heraklion, 700 13 Crete, Greece.

[4]Physics Department, University of Crete, Heraklion, 710 03 Crete, Greece.

∗Corresponding Author: Tel.: +91 1905 267040; Fax: +91 1905 237924

E-mail: suman@iitmandi.ac.in

**Author Contributions**

#A. S. Sarkar and S. Kumari contributed equally.

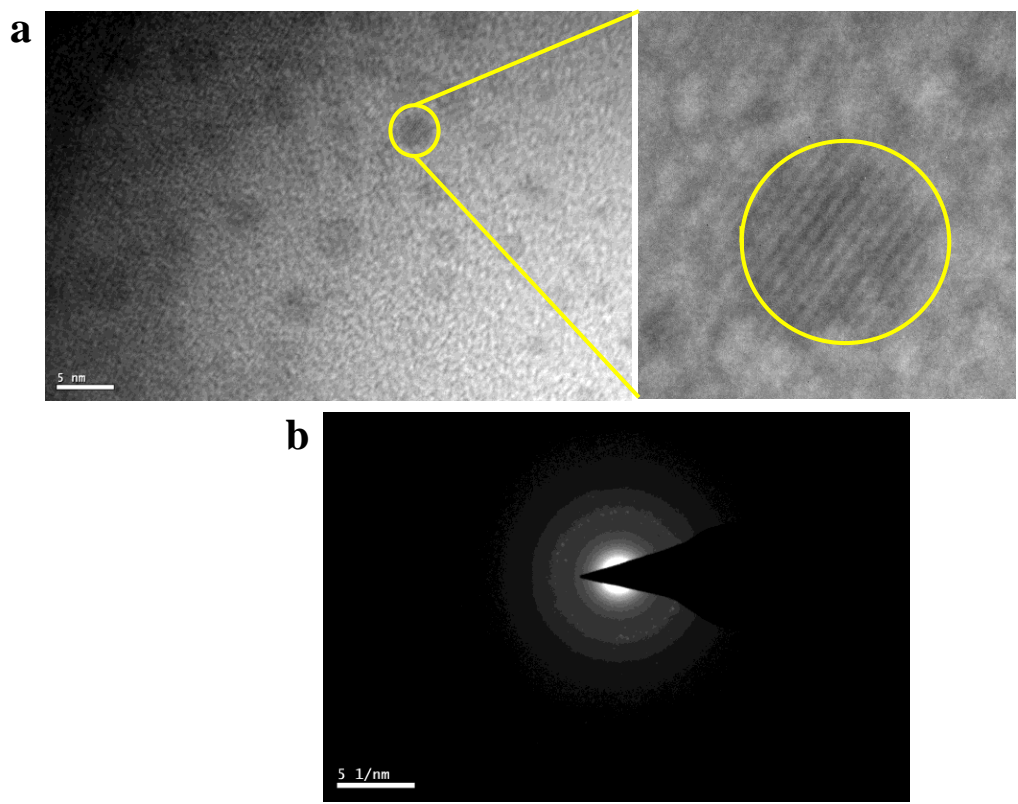

FIG. S1. **a** High-resolution TEM image of SnS quantum dots (QDs). Right panel: zoomed image of a single SnS QD. **b** SAED pattern of the SnS QD.

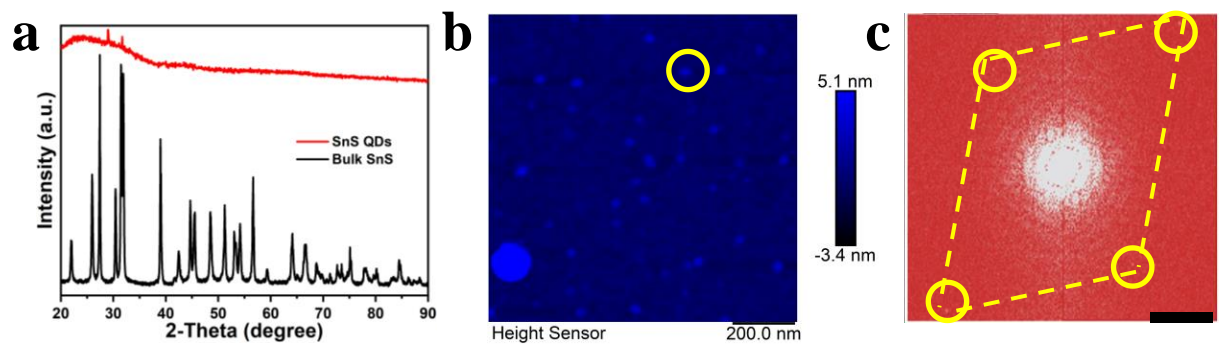

FIG. S2. XRD and AFM images of SnS QDs. **a** X-ray diffraction pattern of bulk SnS and SnS QDs. **b** 2D topographic image, **c** First Fourier transform of the original lattice resolution image encircled in **b**. Scale bar: 2.25 nm$^{-1}$.

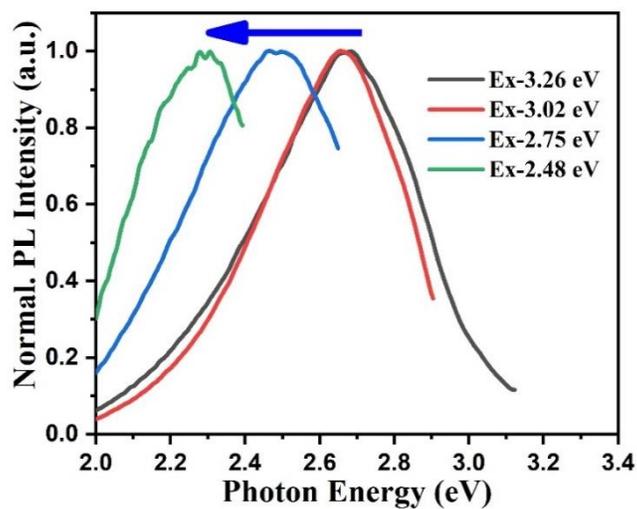

FIG. S3. Normalized excitation dependent PL spectra of SnS QDs.

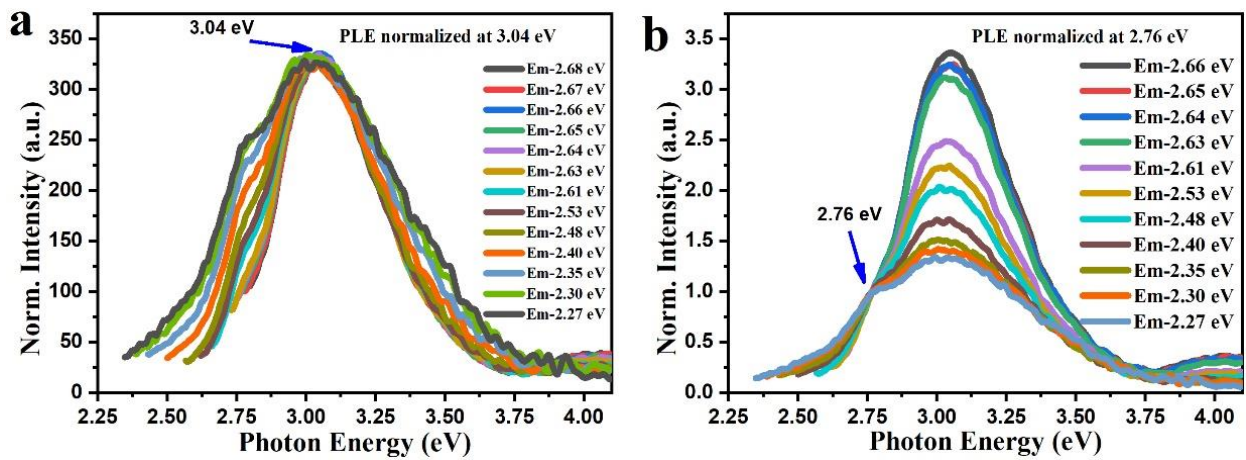

FIG. S4. Normalized PLE spectra. Normalized **a** at 3.04 eV (407 nm), **b** at 2.76 eV (448 nm).